\documentstyle[12pt]{article}
\begin{document}
\begin{titlepage}
\title{Resonance Model of $\pi \Delta \rightarrow Y K$
for Kaon Production in Heavy Ion Collisions}
\author{K. Tsushima\thanks{Supported by DFG under contract No.
Fa 67/14-1},
S.W. Huang\thanks{Supported by GSI under contract No. 06 T\"U 736},
and Amand Faessler\\ \\
Institut f\"ur Theoretische Physik, Universit\"at T\"ubingen\\
Auf der Morgenstelle 14, D-72076 T\"ubingen, F. R. Germany}
\date{}
\maketitle
\begin{abstract}
The elementary production cross sections
$\pi \Delta \rightarrow Y K$ $(Y=\Sigma,\,\, \Lambda)$
and $\pi N \rightarrow Y K$ are needed to describe
kaon production in heavy ion collisions.
The $\pi N \rightarrow Y K$ reactions were studied previously by
a resonance model.
The model can explain the experimental data quite well \cite{tsu}.
In this article, the total cross sections
$\pi \Delta \rightarrow Y K$
at intermediate energies (from the kaon production threshold to
3 GeV of $\pi \Delta$ center-of-mass energy) are calculated
for the first time using the same resonance model.
The resonances,
$N(1710)\,I(J^P) = \frac{1}{2}(\frac{1}{2}^+)$ and
$N(1720)\, \frac{1}{2} (\frac{3}{2}^+)$ for the $\pi \Delta
\rightarrow \Sigma K$ reactions, and
$N(1650)\, \frac{1}{2} (\frac{1}{2}^-)$,
$N(1710)\, \frac{1}{2} (\frac{1}{2}^+)$
and $N(1720)\, \frac{1}{2} (\frac{3}{2}^+)$
for the $\pi \Delta \rightarrow \Lambda K$ reactions
are taken into account coherently as the intermediate states
in the calculations. Also t-channel $K^*(892) \frac{1}{2}(1^-)$
vector meson exchange is included.
The results show that $K^*(892)$ exchange is neglegible
for the $\pi \Delta \rightarrow \Sigma K$ reactions, whereas this meson
does not contribute to the $\pi \Delta \rightarrow \Lambda K$ reactions.
Furthemore, the $\pi \Delta \rightarrow Y K$
contributions to kaon production in heavy ion collisions are not
only non-neglegible but also very different from the $\pi N
\rightarrow Y K$ reactions. An argument valid for
$\pi N \rightarrow Y K$ cannot be extended to
$\pi \Delta \rightarrow Y K$ reactions.
Therefore, cross sections for $\pi \Delta \rightarrow Y K$
including correctly the different isospins must be calculated to
be included in simulation codes for kaon production in heavy
ion collisions, where no experimental data are available.
Parametrizations of the total cross sections
$\pi \Delta \rightarrow Y K$ for
kaon production in heavy ion collisions are given
based on this work.
\end{abstract}
\end{titlepage}%
%
Due to a long mean free path the $K^+$ meson is a good probe
for highly compressed nuclear matter formed in
heavy ion collisions \cite{sch}. $K^+$ production is sensitive
to the nuclear equation of state (EOS) \cite{aic}.
Thus many studies of $K^+$ production in heavy ion collisions have
been performed by theoretically and experimentally \cite{sch}-\cite{cug}.

However, in most theoretical studies of kaon production
by either the Vlasov-Uehling-Uhlenbeck
approach (VUU) \cite{vuu}, or by ``quantum'' molecular dynamics (QMD)
\cite{qmd,li}, the kaon elementry production cross sections
parametrized by J. Randrup and C. M. Ko \cite{ran}, and
by J. Cugnon and R. M. Lombard \cite{cug} have been used.
In these works, the amplitudes relevant for the
elementary kaon production cross sections are not calculated.
Due to the lack of data an average over isospin projections is
used.

Motivated by this fact, we presented for the first time
parametrizations of the total cross sections
$\pi N \rightarrow \Sigma K$
based on theoretical calculations \cite{tsu}.
The processes $\pi N \rightarrow Y K$ $(Y = \Sigma, \Lambda)$
are the so-called secondary processes in
heavy-ion collisions, which are known to give about a
30 \% contribution to kaon production in heavy ion collisions \cite{hal}.
It turned out that the model can explain the
total cross sections $\pi N \rightarrow \Sigma K$ quite well \cite{tsu}.

On the other hand, it was shown by W. Ehehalt. et al. \cite{ehe}
that the $\Delta$'s in the central cell of the
heavy ion collisions make up about 25 \% of the baryons,
whereas the nucleons represent about 60 \% to 70 \%.
This relatively high amount of the $\Delta$'s
in heavy ion collisions is ascribed to the suppression
of the decay $\Delta \rightarrow \pi N$ by Pauli blocking.
This implies that the processes $\pi \Delta \rightarrow Y K$
must also be taken into account for kaon production in heavy-ion
collisions.

However, no experimental data are available for the
$\pi \Delta \rightarrow Y K$ reactions needed to simulate kaon production.

In this article, we will give parametrizations of the total
cross sections  $\pi \Delta \rightarrow Y K$
for the first time based on theoretical calculations.

According to the compilation of the ``Review of Particle
Properties'' \cite{par,paro}, one can select the resonances
$N(1710)\, I(J^P) = \frac{1}{2}(\frac{1}{2}^+)$ and
$N(1720)\, \frac{1}{2}(\frac{3}{2}^+)$ for
$\pi \Delta \rightarrow \Sigma K$,
and the resonances $N(1650)\, \frac{1}{2}(\frac{1}{2}^-),\,\,
N(1710)\, \frac{1}{2}(\frac{1}{2}^+),\,\,$ and
$N(1720)\, \frac{1}{2}(\frac{3}{2}^+)$ for
$\pi \Delta \rightarrow \Lambda K$, as intermediate
states giving the main contributions. For t-channel $K^*$ meson
exhanges, we consider the lightest $K^*(892)\frac{1}{2}(1^-)$
for $\pi \Delta \rightarrow \Sigma K$, where no isospin $I =
1/2$ $K^*$ meson contributes to $\pi \Delta \rightarrow \Lambda K$.

Effective interaction Lagrangians relevant for the
$\pi \Delta \rightarrow Y K$ reactions depicted in fig. 1
are used:
\begin{equation}
{\cal L}_{\pi \Delta N(1650)} =
i \frac{g_{\pi \Delta N(1650)}}{m_\pi}
\left( \bar{N}(1650) \gamma_5 {\overrightarrow{\cal I}}^\dagger
\Delta^\mu \cdot \partial_\mu \vec\phi
+ \bar{\Delta}^\mu \overrightarrow{\cal I} \gamma_5 N(1650) \cdot
\partial_\mu \vec\phi\, \right), \label{lagfirst}
\end{equation}
\begin{equation}
{\cal L}_{\pi \Delta N(1710)} =
\frac{g_{\pi \Delta N(1710)}}{m_\pi}
\left( \bar{N}(1710) {\overrightarrow{\cal I}}^\dagger
\Delta^\mu \cdot \partial_\mu \vec\phi
+ \bar{\Delta}^\mu \overrightarrow{\cal I} N(1710) \cdot
\partial_\mu \vec\phi\, \right),
\end{equation}
\begin{equation}
{\cal L}_{\pi \Delta N(1720)} =
- i g_{\pi \Delta N(1720)}
\left( \bar{N}^\mu(1720) \gamma_5
{\overrightarrow{\cal I}}^\dagger \Delta_\mu  \cdot \vec\phi
+ \bar{\Delta}_\mu \overrightarrow{\cal I} \gamma_5 N^\mu(1720)
\cdot \vec\phi \, \right),
\end{equation}
\begin{equation}
{\cal L}_{K \Sigma N(1710)} =
-ig_{K \Sigma N(1710)}
\left( \bar{N}(1710) \gamma_5 \vec\tau \cdot \overrightarrow\Sigma K
+ \bar{K} \overrightarrow{\bar \Sigma} \cdot \vec\tau
\gamma_5 N(1710) \right),
\end{equation}
\begin{equation}
{\cal L}_{K \Sigma N(1720)} =
\frac{g_{K \Sigma N(1720)}}{m_K}
\left( \bar{N}^\mu(1720) \vec\tau \cdot \overrightarrow\Sigma
\partial_\mu K + (\partial_\mu \bar{K}) \overrightarrow{\bar \Sigma}
\cdot \vec\tau N^\mu(1720) \right),
\end{equation}
\begin{equation}
{\cal L}_{K^*(892) \Sigma \Delta} = - i g_{K^*(892) \Sigma \Delta}
\left( \bar{K}^*_\mu(892) {\overrightarrow{\cal I}}^\dagger
\gamma_5 \Delta^\mu
+ \bar{\Delta}^\mu \gamma_5 \overrightarrow{\cal I}
K^*_\mu(892) \, \right),
\end{equation}
\begin{equation}
{\cal L}_{K^*(892) K \pi} = i f_{K^*(892) K \pi}
\left( \bar{K} \vec\tau K^*_\mu(892) \cdot \partial^\mu \vec\phi
- (\partial^\mu \bar{K}) \vec\tau K^*_\mu(892) \cdot \vec\phi
\, \right) + {\rm h. c.} ,
\end{equation}
\begin{equation}
{\cal L}_{K \Lambda N(1650)} =
- g_{K \Lambda N(1650)} \left( \bar{N}(1650) \Lambda K
+ \bar{K} \bar{\Lambda} N(1650) \right),
\end{equation}
\begin{equation}
{\cal L}_{K \Lambda N(1710)} =
-ig_{K \Lambda N(1710)}
\left( \bar{N}(1710) \gamma_5 \Lambda K
+ \bar{K} \bar{\Lambda} \gamma_5 N(1710) \right),
\end{equation}
\begin{equation}
{\cal L}_{K \Lambda N(1720)} =
\frac{g_{K \Lambda N(1720)}}{m_K}
\left( \bar{N}^\mu(1720) \Lambda
\partial_\mu K + (\partial_\mu \bar{K}) \bar{\Lambda}
N^\mu(1720) \right), \label{laglast}
\end{equation}
where spin 3/2 Rarita-Schwinger particle fields
$\psi^\mu = N^\mu(1720)$ and $\Delta^\mu(1920)$ with mass $m$
satisfy the set of equations \cite{tak},
\begin{eqnarray}
 & ( i \gamma \cdot \partial - m ) \psi^\mu = 0, \\
 & \gamma_\mu \psi^\mu = 0, \\
 & \partial_\mu \psi^\mu = 0.
\end{eqnarray}
$\vec{\cal I}$ is the transition operator defined by
\begin{eqnarray*}
\overrightarrow{\cal I}_{Mm} &= \displaystyle{\sum_{\ell=\pm1,0}}
(1 \ell \frac{1}{2} m | \frac{3}{2} M)
\hat{e}^*_{\ell}\,\,,
\end{eqnarray*}
and $\vec \tau$ are the Pauli matrices.
$N, \Delta^\mu, N(1650), N(1710)$ and $N^\mu(1720)$
stand for the fields of $N(938)$, $\Delta(1232)$,
$N(1650)$, $N(1710)$ and $N(1720)$ resonances.
They are expressed by $\bar{N} = \left( \bar{p},\,\, \bar{n} \right)$,
similarly to the nucleon resonances, and
$\bar{\Delta}_\mu = \left( \bar{\Delta}_\mu^{++},\,\, \bar{\Delta}_\mu^+,\,\,
\bar{\Delta}_\mu^0,\,\, \bar{\Delta}_\mu^- \right)$.
The other field operators appearing in the Lagrangians are
related to the physical representations as follows:
$K^T = \left( K^+,\, K^0 \right),\,\,
\bar{K} = \left( K^-,\, \bar{K^0} \right),\,\,
K^*_\mu(892)^T = \left( K^*_\mu(892)^+,\, K^*_\mu(892)^0 \right),\,\,
\newline \bar{K}^*_\mu(892) = \left( K^*_\mu(892)^-,\,
\bar{K}^*_\mu(892)^0 \right),\,\,
\pi^{\pm} = \frac{1}{\sqrt{2}} (\phi_1 \mp i \phi_2),\,\,
\pi^0 = \phi_3,\,\,
\Sigma^{\pm} = \frac{1}{\sqrt{2}} (\Sigma_1 \mp i \Sigma_2),\,\,
\Sigma^0 = \Sigma_3$, where the superscript
$T$ means the transposition operation.
Here the pseudoscalar meson field operators
are defined annihilating (creating) physical particle
(anti-particle) states. $SU(2)$ isospin symmetry is
assumed for each doublet or multiplet.

We use for the propagators $S_F(p)$ of the spin 1/2 and
$G^{\mu \nu}(p)$ of the spin 3/2 resonances,
\begin{equation}
S_F(p) = \frac{\gamma \cdot p + m}{p^2 - m^2 + im\Gamma^{full}}\,,
\end{equation}
\begin{equation}
G^{\mu \nu}(p) = \frac{P^{\mu \nu}(p)}{p^2 - m^2 +
im\Gamma^{full}}\,,
\end{equation}
with
\begin{equation}
P^{\mu \nu}(p) = - (\gamma \cdot p + m)
\left[ g^{\mu \nu} - \frac{1}{3} \gamma^\mu \gamma^\nu
- \frac{1}{3 m}( \gamma^\mu p^\nu - \gamma^\nu p^\mu)
- \frac{2}{3 m^2} p^\mu p^\nu \right], \label{pmunu}
\end{equation}
\newline
where $m$ and $\Gamma^{full}$ stand for the mass and the full decay
width of the corresponding resonance. In a previous study
\cite{tsu}, we investigated the difference between the results using
the energy dependent decay widths and the results using the energy
independent decay widths.
The two results show that the difference is not significant.
Thus, in order to avoid introducing extra ambiguities,
we use here the energy independent full decay widths for the
propagators of the resonances, since the form
of energy dependent decay width is not uniquely established.
The different ways to introduce decay widths
into the propagators of spin 1/2 and spin 3/2 resonances,
and also other problems concerning the propagators of spin 3/2
particles are discussed by Benmerouche et al. in detail \cite{ben}.

Without factors arising from the isospin structure,
we define the amplitudes ${\cal M}_a, {\cal M}_b$,
${\cal M}_c$ and ${\cal M}_d$ corresponding to each
diagram (a), (b), (c) and (d) given in fig. 1 as follows:
\begin{equation}
{\cal M}_a = \frac{- g_{\pi \Delta N(1650)}g_{K \Lambda N(1650)}}{m_\pi}
\,\,\frac{\,{p_\pi}_{\mu}\, \bar{u}_\Lambda (p_\Lambda)\,
(\gamma \cdot p + m_{N(1710)})\,\gamma_5\, u_\Delta^\mu(p_\Delta)}
{p^2 - m_{N(1650)}^2 + i m_{N(1650)}
\Gamma_{N(1650)}^{full}} , \label{ma}
\end{equation}
\begin{equation}
{\cal M}_b = \frac{- g_{\pi \Delta N(1710)}g_{K Y N(1710)}}{m_\pi}
\,\,\frac{\,{p_\pi}_{\mu}\, \bar{u}_Y (p_Y)\, \gamma_5\,
(\gamma \cdot p + m_{N(1710)})\, u_\Delta^\mu(p_\Delta)}
{p^2 - m_{N(1710)}^2 + i m_{N(1710)}
\Gamma_{N(1710)}^{full}} , \label{mb}
\end{equation}
\begin{equation}
{\cal M}_c = \frac{g_{\pi \Delta N(1720)}g_{K Y N(1720)}}{m_K}
\,\,\frac{\,{p_K}_{\mu}\, \bar{u}_Y (p_Y)\,
P_{N(1720)}^{\mu \nu}(p)\, \gamma_5\,
u_{\Delta \nu}(p_\Delta)}
{p^2 - m_{N(1720)}^2 + i m_{N(1720)}
\Gamma_{N(1720)}^{full}} , \label{mc}
\end{equation}
\begin{equation}
{\cal M}_d = \frac{i f_{K^*(892) K \pi}g_{K^*(892) \Sigma \Delta}}
{(p_\Sigma - p_\Delta)^2 - m^2_{K^*(892)}}
\,\, \bar{u}_\Sigma (p_\Sigma) \gamma_5 u_\Delta^\mu(p_\Delta)
(p_\pi + p_K)^\nu
\left( g_{\mu \nu}
- \frac{(p_\Sigma - p_\Delta)_\mu (p_\Sigma - p_\Delta)_\nu}
{m^2_{K^*(892)}} \right), \label{md}
\end{equation}
where $u_\Delta^\mu(p_\Delta)$, $u_\Lambda(p_\Lambda)$ and
$u_Y(p_Y)$ are the (vector-) spinors of the $\Delta$, the $\Lambda$
and in general for the hyperons $(Y = \Lambda, \Sigma)$ with the momenta
$p_\Delta$, $p_\Lambda$ and $p_Y$, respectively.
Note that the $N(1650)$ resonance
contributes to $\pi \Delta \rightarrow \Lambda K$
but not to $\pi \Delta \rightarrow \Sigma K$.
On the other side, t-channel $K^*(892)$ exchange contributes to
the $\pi \Delta \rightarrow \Sigma K$ but not to
the $\pi \Delta \rightarrow \Lambda K$ reactions.

Then each amplitude of the $\pi \Delta \rightarrow Y K$
reactions is given by:
\\ \\
\noindent
{\bf For the $\pi \Delta \rightarrow \Sigma K$ reactions:}
\begin{eqnarray}
- ({\cal M}_b+{\cal M}_c)\qquad {\rm for}
&\pi^- \Delta^{++} \rightarrow \Sigma^0 K^+ \quad
{\rm and}\quad \pi^+ \Delta^- \rightarrow \Sigma^0 K^0, \label{ampfirst}\\
\mp \sqrt{\frac{2}{3}} ({\cal M}_b+{\cal M}_c)\qquad {\rm for}
&\pi^- \Delta^+ \rightarrow  \Sigma^- K^+ \quad
{\rm and}\quad \pi^+ \Delta^0 \rightarrow \Sigma^+ K^0,\\
- {\cal M}_d \qquad {\rm for}
&\pi^0 \Delta^{++} \rightarrow  \Sigma^+ K^+ \quad
{\rm and}\quad \pi^0 \Delta^- \rightarrow \Sigma^- K^0,\\
\pm \sqrt{\frac{2}{3}} ({\cal M}_b+{\cal M}_c+{\cal M}_d)\qquad {\rm for}
&\pi^0 \Delta^+ \rightarrow \Sigma^0 K^+ \quad
{\rm and}\quad \pi^0 \Delta^0 \rightarrow \Sigma^0 K^0,\\
\frac{1}{\sqrt{3}} (2{\cal M}_b+2{\cal M}_c+{\cal M}_d)\qquad {\rm for}
&\pi^0 \Delta^0 \rightarrow \Sigma^- K^+ \quad
{\rm and}\quad \pi^0 \Delta^+ \rightarrow \Sigma^+ K^0,\\
\mp \sqrt{\frac{2}{3}}{\cal M}_d\qquad {\rm for}
&\pi^+ \Delta^+ \rightarrow \Sigma^+ K^+ \quad
{\rm and}\quad \pi^- \Delta^0 \rightarrow \Sigma^- K^0,\\
\frac{1}{\sqrt{3}} ({\cal M}_b+{\cal M}_c+2{\cal M}_d)\qquad {\rm for}
&\pi^+ \Delta^0 \rightarrow \Sigma^0 K^+ \quad
{\rm and}\quad \pi^- \Delta^+ \rightarrow \Sigma^0 K^0,\\
\pm \sqrt{2} ({\cal M}_b+{\cal M}_c+{\cal M}_d)\qquad {\rm for}
&\pi^+ \Delta^- \rightarrow  \Sigma^- K^+ \quad
{\rm and}\quad \pi^- \Delta^{++} \rightarrow \Sigma^+ K^0,
\label{amplast}
\end{eqnarray}
\\
\noindent
{\bf For the $\pi \Delta \rightarrow \Lambda K$ reactions:}
\begin{eqnarray}
\mp ({\cal M}_a+{\cal M}_b+{\cal M}_c)\qquad {\rm for}
&\pi^- \Delta^{++} \rightarrow  \Lambda K^+ \quad
{\rm and}\quad \pi^+ \Delta^{-} \rightarrow \Lambda K^0,\\
\sqrt{\frac{2}{3}} ({\cal M}_a+{\cal M}_b+{\cal M}_c)\qquad {\rm for}
&\pi^0 \Delta^+ \rightarrow  \Lambda K^+ \quad
{\rm and}\quad \pi^0 \Delta^0 \rightarrow \Lambda K^0,\\
\pm \frac{1}{\sqrt{3}} ({\cal M}_a+{\cal M}_b+{\cal M}_c)\qquad {\rm for}
&\pi^+ \Delta^0 \rightarrow  \Lambda K^+ \quad
{\rm and}\quad \pi^- \Delta^+ \rightarrow \Lambda K^0, \label{ampla}
\end{eqnarray}
where the upper and lower signs in front of the amplitudes
should be assigned to the $K^+$ and $K^0$ channels, respectively.

Next we need to determine the coupling constants appearing
in the Lagrangians eqs. (\ref{lagfirst}) - (\ref{laglast}).
In order to detemine the coupling constants and to perform the
calculations, form factors (denoted by $F(q)$ and
$F_{K^*(892) K \pi}(q)$ below)
are introduced which represent the finite size of the hadrons.
These form factors must be multiplied to each vertex of the interactions.
Thus, the coupling constants are obtained from the branching
ratios in the rest frame of the resonances:
$$
\Gamma(N(1650) \rightarrow \Delta \pi) =
2 \frac{g^2_{\pi \Delta N(1650)}
F^2(q(m_{N(1650)},m_\Delta,m_\pi))}{6\pi}\hspace{10em}
$$
\begin{equation}
\hspace{16em}\cdot
\frac{m_{N(1650)} (E_\Delta - m_\Delta)}{m_\pi^2 m_\Delta^2}
q^3(m_{N(1650)},m_\Delta,m_\pi),
\end{equation}
$$
\Gamma(N(1710) \rightarrow \Delta \pi) =
2 \frac{g^2_{\pi \Delta N(1710)}
F^2(q(m_{N(1710)},m_\Delta,m_\pi))}{6\pi}\hspace{10em}
$$
\begin{equation}
\hspace{16em}\cdot
\frac{m_{N(1710)} (E_\Delta + m_\Delta)}{m_\pi^2 m_\Delta^2}
q^3(m_{N(1710)},m_\Delta,m_\pi),
\end{equation}
$$
\Gamma(N(1720) \rightarrow \Delta \pi) =
2 \frac{g^2_{\pi \Delta N(1720)}
F^2(q(m_{N(1720)},m_\Delta,m_\pi))}{36\pi}
q(m_{N(1720)},m_\Delta,m_\pi)\hspace{2em}
$$
\begin{equation}
\hspace{14em}
\cdot (\frac{m_\Delta}{m_{N(1720)}}) \left[\,
(\frac{E_\Delta}{m_\Delta}) - 1\,\right]
\left[\,2(\frac{E_\Delta}{m_\Delta})^2 -
2(\frac{E_\Delta}{m_\Delta}) + 5\,\right] ,
\end{equation}
$$
\Gamma(N(1650) \rightarrow \Lambda K) =
\frac{g^2_{K \Lambda N(1650)} F^2(q(m_{N(1650)},m_\Lambda,m_K))}
{4\pi}\hspace{10em}
$$
\begin{equation}
\hspace{16em}\cdot \frac{(E_\Lambda + m_\Lambda)}{m_{N(1650)}}
q(m_{N(1650)},m_\Lambda,m_K),
\end{equation}
$$
\Gamma(N(1710) \rightarrow Y K) =
D \frac{g^2_{K Y N(1710)} F^2(q(m_{N(1710)},m_Y,m_K))}
{4\pi}\hspace{10em}
$$
\begin{equation}
\hspace{16em}\cdot \frac{(E_Y - m_Y)}{m_{N(1710)}}
q(m_{N(1710)},m_Y,m_K), \label{d1}
\end{equation}
\newpage
$$
\Gamma(N(1720) \rightarrow Y K) =
D \frac{g^2_{K Y N(1720)} F^2(q(m_{N(1720)},m_Y,m_K))}
{12\pi}\hspace{9em}
$$
\begin{equation}
\hspace{16em}\cdot \frac{(E_Y + m_Y)}{m_{N(1720)}m_K^2}
q^3(m_{N(1720)},m_Y,m_K), \label{d2}
\end{equation}
$$
\Gamma(K^*(892) \rightarrow K \pi) =
3 \frac{f^2_{K^*(892) K \pi} F_{K^*(892) K \pi}^2(q(m_{K^*(892)},m_K,m_\pi))}
{4\pi} \hspace{10em}
$$
\begin{equation}
\hspace{16em} \cdot \frac{2}{3 m^2_{K^*(892)}}
q^3(m_{K^*(892)},m_K,m_\pi),
\end{equation}
with
\begin{equation}
F(q) = \frac{\Lambda_C^2}{\Lambda_C^2 + q^2}, \qquad
F_{K^*(892) K \pi}(q) = C q \exp \left( - \beta q^2 \right),
\label{form}
\end{equation}
\begin{equation}
q(x,m_B,m_P) = \frac{1}{2 x}
\left[ (x^2 - (m_B + m_P)^2)\,(x^2 - (m_B - m_P)^2) \right]^{1/2},
\label{qdeff}
\end{equation}
\newline
where $B^*$, $B$ and $P$ in $q(m_{B^*},m_B,m_P)$
stand for the relevant resonance, the baryon and the pseudoscalar meson,
respectively.
$q = q(m_{B^*},m_B,m_P)$ satisfies $q = |\vec{p}_B|$ with
$\vec{p}_B = - \vec{p_P}$ and
$E_B = \sqrt{m_B^2 + \vec{p}\hspace{1mm}^2_B}$.
$F(q)$ is the form factor with the cut-off parameter
$\Lambda_C$, and $Y$ stands for either the $\Sigma$ or the $\Lambda$.
The constant $D$ in eqs. (\ref{d1}) and (\ref{d2}) should be assigned
to $D = 3$ for $Y = \Sigma$
and $D = 1$ for $Y = \Lambda$, respectively.
The $K^*(892) K \pi$ vertex form factor is taken from ref.
\cite{gob}.
The calculated coupling constants and the experimental data used
to determine them are given in Tables 1 and 2.
We use the value of the $g_{K^*(892) \Sigma N}$ for the
$g_{K^*(892) \Sigma \Delta}$, which obtained by fitting the $\pi^+ p
\rightarrow \Sigma^+ K^+$ channel in the previous study \cite{tsu}.
Note that in this case an extra factor $\sqrt{3}$ must be
included due to the different normalization of the operator
in isospin space $\vec\tau$ and $\overrightarrow{\cal I}$.
In evaluating
the cross sections in the center-of-mass frame of the $\pi \Delta$ system,
each coupling constants
$g_{PBB^*}$, $g_{K^*(892) \Sigma \Delta}$ and $f_{K^*(892) K \pi}$
appearing in eqs. (\ref{ma}) - (\ref{md})
must be replaced by
$g_{PBB^*} \rightarrow g_{PBB^*}F(q(\sqrt{s},m_B,m_P))$,
$g_{K^*(892) \Sigma \Delta} \rightarrow
g_{K^*(892) \Sigma \Delta}F((\vec{q}_f - \vec{q}_i))$
and $f_{K^*(892) K \pi} \rightarrow f_{K^*(892) K
\pi}F_{K^*(892) K \pi}( \frac{1}{2}(\vec{q}_f - \vec{q}_i) )$,
where $s$ is the Mandelstam variable,
$q(\sqrt{s},m_B,m_P) = |\vec{p_B}|$,
$\vec{p}_B = - \vec{p}_P$, $|\vec{q}_f| = q(\sqrt{s},m_\Sigma,m_K)$
and $|\vec{q}_i| = q(\sqrt{s},m_\Delta,m_\pi)$.
We use the same value for the cut-off parameter $\Lambda_C$
fixed by the previous study of the
$\pi N \rightarrow \Sigma K$ reactions \cite{tsu}, i.e.
$\Lambda_C = 0.8$ GeV for all resonances considerd here,
$\Lambda_C = 1.2$ GeV for the $K^*(892) \Sigma \Delta$ vertex.
The parameters $C$ and $\beta$ in $F_{K^*(892) K \pi}$ are
$C = 2.72$ fm and $\beta = 8.88 \times 10^{-3}$ fm$^2$ used in ref.
\cite{gob}.

Hereafter, we will discuss the $K^+$ production channels only.
Corresponding arguments for the $K^0$ production channels should be
valid as seen from eqs. (\ref{ampfirst}) - (\ref{amplast}).

The calculated total cross sections are displayed in figs. 2
(a), 2 (b) and 3, corresponding to the
$\pi^- \Delta^{++} \rightarrow \Sigma^0 K^+$,
the $\pi^+ \Delta^- \rightarrow \Sigma^- K^+$ and
the $\pi^- \Delta^{++} \rightarrow \Lambda K^+$ reactions, respectively.

We discuss the results of the
$\pi \Delta \rightarrow \Sigma K$
reactions given in figs. 2 (a) and 2 (b) first.
The solid lines show them without
interference terms and the dashed lines with interference terms.
There are four possibilities for the sign combination of the interference
terms. The largest and the smallest results for each relevant
channel are shown among the four different sign combinations of
the coupling constants. It turned out that the $K^*(892)$
exchange contribution is neglegible for the $\pi \Delta
\rightarrow \Sigma K$ reactions. Typically a square of the
$K^*(892)$ exhange amplitude $|{\cal M}_d|^2$ is more
than one order of magnitudes smaller compared to the other
contributions. (See eqs. (\ref{ampfirst}) - (\ref{amplast}).)
This is different from $\pi N \rightarrow Y K$ reactions where
$K^*(892)$ exchange gives the same order for the contributions as
other resonances.

According to our previous study \cite{tsu}, total cross sections
have in their peaks for $\pi N \rightarrow \Sigma K$ about
0.2 to 0.4 mb, except for $\pi^+ p \rightarrow \Sigma^+ K^+$,
where the experimental data show about 0.75 mb. It is clear that
the total cross sections for
$\pi \Delta \rightarrow \Sigma K$ are in their peaks of the same order
of those for $\pi N \rightarrow \Sigma K$. Furthermore, the energy
dependence is also different from $\pi N \rightarrow \Sigma K$.
Thus, the argument valid for $\pi N \rightarrow \Sigma K$
reactions cannot be extended to the $\pi \Delta \rightarrow
\Sigma K$ reactions. Thus difference exists not only for the absolute
value but also the energy dependence of the total cross sections.

Next, we discuss the results of
$\pi^- \Delta^{++} \rightarrow \Lambda K^+$ displayed in fig. 3.
They are also given for three cases:
Without interference terms (the solid line), and with
interefrence terms (the dashed lines). There are four
possibilities for the sign combination of the interference terms.
Again the largest and the smallest results among the four different sign
combinations are shown in fig. 3.
The total cross section at this peak position is of about a 30 \%
of the largest channel for the $\pi N \rightarrow \Lambda K$
reactions, i.e. for $\pi^+ n \rightarrow \Lambda K^+$.
The energy dependence of the total cross sections for
$\pi \Delta \rightarrow \Lambda K$ is rather simillar to that of
$\pi N \rightarrow \Lambda K$.

It should be emphasized again here that, the total cross sections
for $\pi \Delta \rightarrow Y K$ to kaon production are not
small at all. Furthermore, no quantitative argument for
kaon production cross sections $\pi \Delta \rightarrow Y K$
so far has been given based on theoretical calculations,
nor based on experimental data.

Now, we are in a position to give parametrizations of
total cross sections
$\pi^- \Delta^{++} \rightarrow \Sigma^0 K^+$,
$\pi^0 \Delta^0 \rightarrow \Sigma^- K^+$,
$\pi^+ \Delta^0 \rightarrow \Sigma^0 K^+$,
$\pi^+ \Delta^- \rightarrow \Sigma^- K^+$ and
$\pi^- \Delta^{++} \rightarrow \Lambda K^+$ which
are enough to reproduce whole channel parametrizations given in
eqs. (\ref{ampfirst}) - (\ref{ampla}).  The channels only the
$K^*(892)$ exchange gives contribution are omitted since they
are neglegible as mentioned before.
Since the signs of interference terms cannot be fixed by
experimental data, we parametrize the results
obtained without intereference terms (solid lines in the figures
2 (a) to 3). They are:
\\
\\
\noindent{\bf For $\pi \Delta \rightarrow \Sigma K$:}
\begin{equation}
\sigma(\pi^- \Delta^{++} \rightarrow \Sigma^0 K^+)
= \frac{0.004959 (\sqrt{s}-1.688)^{0.7785}}{(\sqrt{s}-1.725)^2+0.008147}
\hspace{0.5cm} {\rm mb},\hspace{11em}
\end{equation}
\begin{equation}
\sigma(\pi^0 \Delta^0 \rightarrow \Sigma^- K^+)
= \frac{0.006964 (\sqrt{s}-1.688)^{0.8140}}{(\sqrt{s}-1.725)^2+0.007713}
\hspace{0.5cm} {\rm mb},\hspace{11em}
\end{equation}
\begin{equation}
\sigma(\pi^+ \Delta^0 \rightarrow \Sigma^0 K^+)
=  \frac{0.002053 (\sqrt{s}-1.688)^{0.9853}}{(\sqrt{s}-1.725)^2+0.005414}
 + \frac{0.3179 (\sqrt{s}-1.688)^{0.9025}}{(\sqrt{s}-2.675)^2+44.88}
\hspace{0.5cm} {\rm mb},
\end{equation}
\begin{equation}
\sigma(\pi^+ \Delta^- \rightarrow \Sigma^- K^+)
= \frac{0.01741 (\sqrt{s}-1.688)^{1.2078}}{(\sqrt{s}-1.725)^2+0.003777}
\hspace{0.5cm} {\rm mb},\hspace{11em}
\end{equation}
\\
{\bf For $\pi \Delta \rightarrow \Lambda K$:}
\begin{equation}
\sigma(\pi^- \Delta^{++} \rightarrow \Lambda K^+)
 = \frac{0.006545 (\sqrt{s}-1.613)^{0.7866}}{(\sqrt{s}-1.720)^2+0.004852}
\hspace{0.5cm} {\rm mb},\hspace{11em}
\end{equation}
where, the parametrizations for
$\sigma(\pi \Delta \rightarrow \Sigma K)$ and
$\sigma(\pi^- \Delta^{++} \rightarrow \Lambda K^+)$
given above should be understood
to be zero below the thresholds $\sqrt{s} \leq 1.688$ GeV and
$\sqrt{s} \leq 1.613$ GeV, respectively.
These parametrizations are useful for codes which simulate
kaon production since no experimental data are available.
In earlier work \cite{tsu} we gave already parametrizations
for the reactions $\pi N \rightarrow \Sigma K$ based on similar
calculations with intermediate resonances and $K^*(892)$ exchanges.

To summarize, we studied the $\pi \Delta \rightarrow Y K$
reactions by a resonance model, and presented for the first
time explicite parametrizations of the energy dependence of
their total cross sections.
It turned out that the t-channel $K^*(892)$ exchange
contribution for $\pi \Delta \rightarrow \Sigma K$ is neglegible.
Furthermore, the contributions of the
$\pi \Delta \rightarrow Y K$ reactions to kaon production in
heavy ion collisions are not only non-neglegible, but also very
different from the $\pi N \rightarrow Y K$ contributions.
Therefore, the $\pi \Delta \rightarrow Y K$ contributions must be
adequatly included into the studies of kaon production in heavy
ion collisions without relying on isospin arguments relating the
cross sections for isospin $I = 1/2$ nucleons and the isospin
$I = 3/2$ $\Delta$'s.
\vspace{2cm}

\noindent {\bf Acknowledgement:} The authors express their
thanks to Prof. K. W. Schmid for providing us a code to a least
square fit for adjusting the parameters.

\newpage
\begin{table}
\caption{Coupling constants for the
$\pi \Delta \rightarrow \Sigma K$ reactions}
\begin{center}
\begin{tabular}{cccccc}
\hline
$B^*$(resonance) &$\Gamma(MeV)$
&$\Gamma_{\Delta \pi}(\%)$  &$g_{B^* \Delta \pi}^2$
&$\Gamma_{\Sigma K}(\%)$    &$g_{B^* \Sigma K}^2$  \\
\hline \\
$N(1710)$     &100 &17.5 &$1.85\times10^{-2}$&6.0 &$4.50\times10^{+1}$\\
$N(1720)$     &150 &10.0 &$1.12\times10^{+1}$&3.5 &3.15\\
\\
\hline
 &$f^2_{K^*(892) K \pi}$&
 &$g^2_{K^*(892) \Sigma \Delta}$ & & \\
\hline
\\
 &$6.89\times10^{-1}$& &$6.08\times10^{-1}$& & \\
 &($\Gamma = 50$ MeV,&$\Gamma_{K \pi} = 100 \%$)& & & \\
\\ \hline
\end{tabular}
\end{center}
\end{table}
\vspace{2cm}
\begin{table}
\caption{Coupling constants for the
$\pi \Delta \rightarrow \Lambda K$ reactions}
\begin{center}
\begin{tabular}{cccccc}
\hline
$B^*$(resonance) &$\Gamma(MeV)$
&$\Gamma_{\Delta \pi}(\%)$   &$g_{B^* \Delta \pi}^2$
&$\Gamma_{\Lambda K}(\%)$    &$g_{B^* \Lambda K}^2$  \\
\hline \\
$N(1650)$     &150 & 5.0 &$6.56\times10^{-1}$& 7.0 &$6.40\times10^{-1}$\\
$N(1710)$     &100 &17.5 &$1.85\times10^{-2}$&15.0 &$4.74\times10^{+1}$\\
$N(1720)$     &150 &10.0 &$1.12\times10^{+1}$& 6.5 &3.91\\
\\
\hline
\end{tabular}
\end{center}
\end{table}
\vspace{1.5cm}
\noindent
{\bf Table 1}
\newline
The calculated coupling constants and the experimental data for the $\pi
\Delta \rightarrow \Sigma K$ reactions.
\vspace{1.5cm}

\noindent
{\bf Table 2}
\newline
The calculated coupling constants and the experimental data for the $\pi
\Delta \rightarrow \Lambda K$ reactions.

\newpage
\noindent
{\bf Figure captions}
\vspace{1.5cm}

\noindent
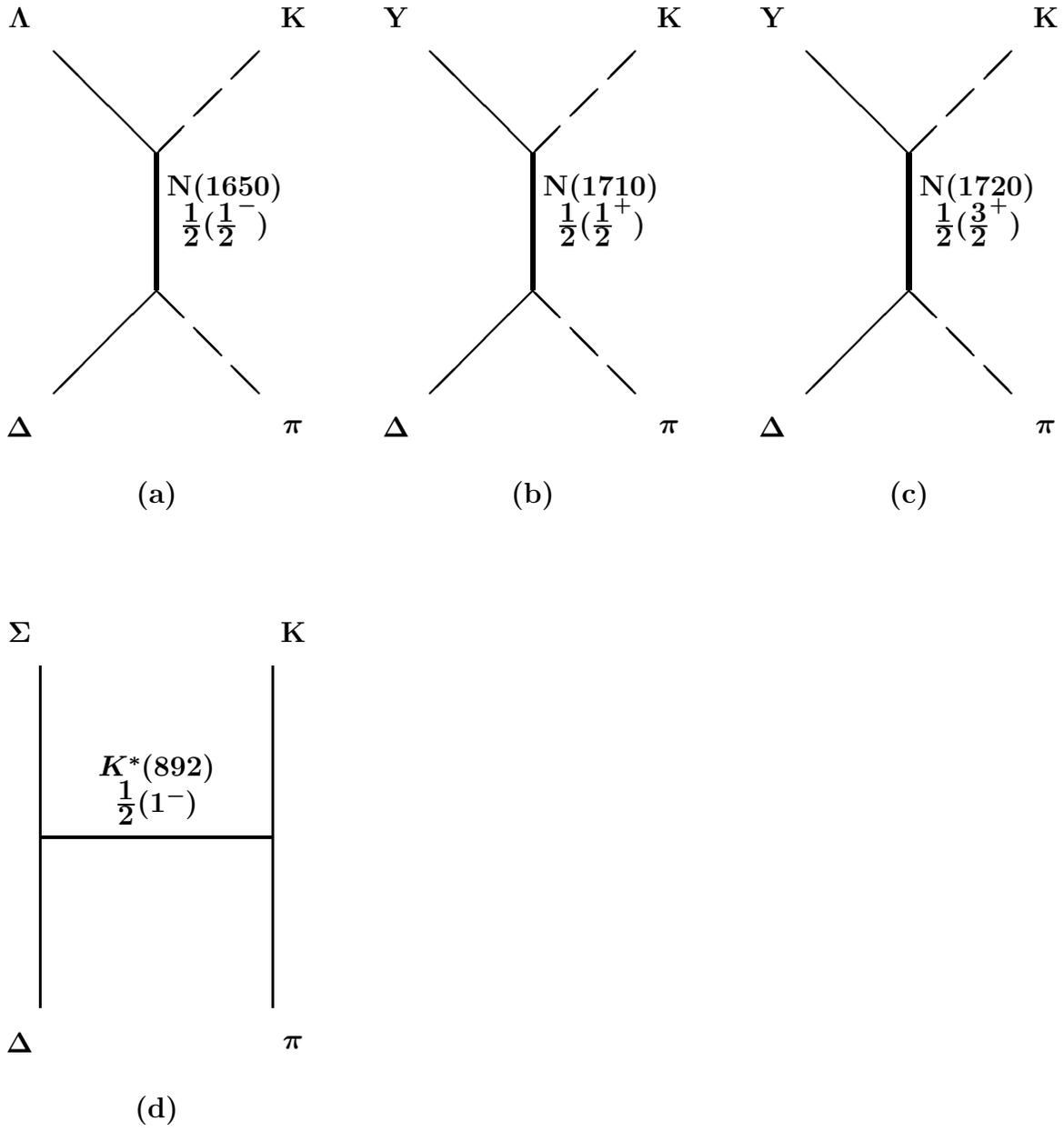
{\bf Fig. 1}
\newline
The preocesses contributing to the $\pi \Delta \rightarrow Y K$
$(Y = \Sigma, \Lambda)$ reactions. The diagrams are corresponding
to the different intermediate resonance states;
$(a): N(1650)\, I(J^P) = \frac{1}{2}(\frac{1}{2}^-)$,
$(b): N(1710)\, \frac{1}{2}(\frac{1}{2}^+)$ and
$(c): N(1720)\, \frac{1}{2}(\frac{3}{2}^+)$, respectively.
\vspace{1cm}

\noindent
{\bf Fig. 2 (a)}
\newline
The calculated total cross sections for the
$\pi^- \Delta^{++} \rightarrow \Sigma^0 K^+$
($\pi^+ \Delta^- \rightarrow \Sigma^0 K^0$) reactions.
The solid line and the dashed lines stand for the results without
and with the inclusion the interference terms, respectively.
Note that the largest and the smallest results are displayed for
the four possibilities arising from the possible signs of the
coupling constants and thus the interference terms.
\vspace{1cm}

\noindent
{\bf Fig. 2 (b)}
\newline
The calculated total cross sections for the
$\pi^+ \Delta^- \rightarrow \Sigma^- K^+$
($\pi^- \Delta^{++} \rightarrow \Sigma^+ K^0$) reactions.
See the caption of fig. 2 (a) for further explanations.
\vspace{1cm}

\noindent
{\bf Fig. 3}
\newline
The calculated total cross sections for the
$\pi^- \Delta^{++} \rightarrow \Lambda K^+$
($\pi^+ \Delta^- \rightarrow \Lambda K^0$) reactions.
See the caption of fig. 2 (a) for further explanations.

\newpage
{\Large \bf Fig. 1}\\
\bf
\boldmath
\setlength{\unitlength}{1cm}
\begin{picture}(11,20) \thicklines
\put(0,11){\makebox(1,1){$\Delta$}}
\put(0,17){\makebox(1,1){$\Lambda$}}
\put(2,10){\makebox(1,1){(a)}}
\put(4,11){\makebox(1,1){$\pi$}}
\put(4,17){\makebox(1,1){K}}
\put(3,14.5){\makebox(1,1){N(1650)}}
\put(3,14){\makebox(1,1)
{$\frac{\displaystyle 1}{\displaystyle 2}
(\frac{\displaystyle 1}{\displaystyle 2}^-)$}}
\put(2.5,13.5){\line(-1,-1){1.5}}
\multiput(2.5,13.5)(0.55,-0.55){3}{\line(1,-1){0.4}}
\put(2.48,13.5){\line(0,1){2}}
\put(2.5,13.5){\line(0,1){2}}
\put(2.52,13.5){\line(0,1){2}}
\put(2.5,15.5){\line(-1,1){1.5}}
\multiput(2.5,15.5)(0.55,0.55){3}{\line(1,1){0.4}}
\put(5.5,11){\makebox(1,1){$\Delta$}}
\put(5.5,17){\makebox(1,1){Y}}
\put(7.5,10){\makebox(1,1){(b)}}
\put(9.5,11){\makebox(1,1){$\pi$}}
\put(9.5,17){\makebox(1,1){K}}
\put(8.5,14.5){\makebox(1,1){N(1710)}}
\put(8.5,14){\makebox(1,1)
{$\frac{\displaystyle 1}{\displaystyle 2}
(\frac{\displaystyle 1}{\displaystyle 2}^+)$}}
\put(8,13.5){\line(-1,-1){1.5}}
\multiput(8,13.5)(0.55,-0.55){3}{\line(1,-1){0.4}}
\put(7.98,13.5){\line(0,1){2}}
\put(8,13.5){\line(0,1){2}}
\put(8.02,13.5){\line(0,1){2}}
\put(8,15.5){\line(-1,1){1.5}}
\multiput(8,15.5)(0.55,0.55){3}{\line(1,1){0.4}}
\put(11,11){\makebox(1,1){$\Delta$}}
\put(11,17){\makebox(1,1){Y}}
\put(13,10){\makebox(1,1){(c)}}
\put(15,11){\makebox(1,1){$\pi$}}
\put(15,17){\makebox(1,1){K}}
\put(14,14.5){\makebox(1,1){N(1720)}}
\put(14,14){\makebox(1,1)
{$\frac{\displaystyle 1}{\displaystyle 2}
(\frac{\displaystyle 3}{\displaystyle 2}^+)$}}
\put(13.5,13.5){\line(-1,-1){1.5}}
\multiput(13.5,13.5)(0.55,-0.55){3}{\line(1,-1){0.4}}
\put(13.48,13.5){\line(0,1){2}}
\put(13.5,13.5){\line(0,1){2}}
\put(13.52,13.5){\line(0,1){2}}
\put(13.5,15.5){\line(-1,1){1.5}}
\multiput(13.5,15.5)(0.55,0.55){3}{\line(1,1){0.4}}
\put(0,2){\makebox(1,1){$\Delta$}}
\put(0,8){\makebox(1,1){$\Sigma$}}
\put(2,1){\makebox(1,1){(d)}}
\put(4,2){\makebox(1,1){$\pi$}}
\put(4,8){\makebox(1,1){K}}
\put(2,6){\makebox(1,1){$K^*$(892)}}
\put(2,5.5){\makebox(1,1)
{$\frac{\displaystyle 1}{\displaystyle 2}({\displaystyle 1}^-)$}}
\put(0.8,3){\line(0,1){5}}
\put(4.2,3){\line(0,1){5}}
\put(0.8,5.5){\line(1,0){3.4}}
\end{picture}
\end{document}